\documentclass[%
 twocolumn,
 amsmath, amssymb,
 aps,
 pre,
]{revtex4-1}

\usepackage{graphicx}
\usepackage{dcolumn}
\usepackage{bm}
\usepackage{multirow}
\usepackage{diagbox}
\usepackage{enumerate}
\renewcommand{\vec}[1]{{\bf #1}}
\newcommand{\braket}[1]{\langle #1  \rangle}
\newcommand{\bbraket}[2]{\langle\langle #1, #2  \rangle\rangle}

\renewcommand{\min}{\text{min}}

\begin{document}

\title{The Tangent Space to the Manifold of Critical Classical Hamiltonians Representable by Tensor Networks} 

\author{Yantao Wu}
\affiliation{
The Department of Physics, Princeton University
}

\date{\today}
\begin{abstract}
We introduce a scheme to perform Monte Carlo Renormalization Group with the coupling constants of the system Hamiltonian encoded in a tensor network.  
With this scheme we compute the tangent space to the manifold of the critical Hamiltonians representable by a tensor network at the nearest-neighbor critical coupling for three models: the two and three dimensional Ising models and the two dimensional three-state Potts model. 
\end{abstract}
\pacs{Valid PACS appear here}
\maketitle

\section{Introduction}
\label{sec:intro}
Since the application of Renormalization Group (RG) to critical phenomena \cite{wilson_rg}, there has been a continuing progress in realizing the RG program non-pertubatively.   
However, to implement a correct RG transformation is not trivial. 
In a proper implementation of renormalization, the theory of RG requires that 
\begin{enumerate}[A.]
\item the non-critical microscopic Hamiltonians flow into trivial fixed-points characteristic of the phases they represent, 
\item different critical microscopic Hamiltonians flow into a unique non-trivial fixed-point in the absence of marginal RG operators.    
\end{enumerate}
In dealing with classical lattice models, two successful numerical implementations of real-space RG are Monte Carlo Renormalization Group (MCRG) \cite{mamcrg,mcrg,mcrg_rc,iMCRG,vmcrg} and a variety of algorithms under the general name tensor network renormalization group (TNRG) \cite{trg,srg,teft,tnr,tnr_algo,loop_tnr,tnr+,gilt,trg+fet}. 
For MCRG, in the various models considered, for example the 2D and 3D Ising models, both requirements have been observed to hold \cite{cmts}. 
For TNRG, however, to the best knowledge of the author, only requirement A has been checked, which even initially presented a challenge to its first example \cite{trg}.      
Various subsequent TNRG algorithms have found different ways to meet requirement A to produce a proper RG flow \cite{teft,tnr,loop_tnr,tnr+,gilt,trg+fet}. 
If requirement B is indeed satisfied in TNRG, one would expect that as a critical microscopic tensor is perturbed along the tangent space of the set of critical Hamiltonians representable by a tensor network, the change in the final renormalized tensor at a sufficiently large RG iteration level should change at most to quadratic order of the perturbation. 
To check this, however, prior knowledge on the behavior of the critical Hamiltonians representable by a tensor network would be necessary.  

In this paper, we describe how MCRG can be performed with coupling constants encoded in a tensor network, and in so doing, determine the tangent space to the set of critical Hamiltonians representable by a tensor network, using a technique recently introduced in MCRG \cite{cmts}.  
The set of all critical Hamiltonians of a system is typically a differentiable manifold, and is called the {\it critical manifold} of the system. 
In the following, we will call the set of critical Hamiltonians representable by a tensor network the {\it tensor network critical manifold} (TNCM), which is a submanifold of the critical manifold.
% This is a very interesting issue which concerns with the deep mystery of universality. 
% We hope that the tangent space to TNCM reported here can help further clarify the renormalization procedure in TNRG algorithms. 
The paper is organized as follows. In Sec. \ref{sec:tn_mcrg}, we introduce the scheme to perform MCRG with tensor networks. In Sec. \ref{sec:cmts}, we review the method to compute the tangent space to the critical manifold with MCRG. In \ref{sec:result}, we report the results on 2D and 3D Ising models and the 2D three-state Potts model. 
In Sec. \ref{sec:discussion}, we conclude.      
\section{Monte Carlo Renormalization Group with Tensor Networks}
\label{sec:tn_mcrg}
\begin{figure}[htb]
\centering
  \centering
  \includegraphics[scale=1.3]{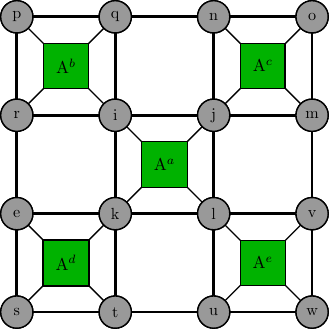}
\caption{Part of a tensor network representing a 2D classical system. $ijkl...$ represent tensor indices.}
\label{fig:2dtn_left}
\end{figure}

We first review the tensor network representation of a classical partition function \cite{trg,tnr_algo}. 
Although the representation is general, for concreteness let us consider the two-dimensional Ising model on a square lattice with the Hamiltonian:   
\begin{equation}
  \label{eq:isingham}
  H(\bm\sigma) = -K\sum_{\braket{x,y}} \sigma_x\sigma_y , 
\end{equation}
where $\braket{x,y}$ denotes nearest-neighbor pairs and $\sigma_x = \pm 1$ on each lattice site.  
$K > 0$ is the coupling constant. 
Its partition function has a tensor network representation \cite{tnr_algo} shown in Fig. \ref{fig:2dtn_left}:  
\begin{equation}
  \label{eq:Z_tensor}
  Z = \sum_{ijk\cdots} A^a_{ijkl} A^b_{pqri} A^c_{nojm} \cdots , 
\end{equation}
where the superscripts, $a, b, c...$, on $A$ denote the distinct tensors located at different positions on the lattice.   
The tensor indices, $ijkl\cdots$, take values $0$ or $1$, labeling a tensor of bond dimension $\chi = 2$. 
%----------------%
\begin{figure}[htb]
\centering
\centering
\includegraphics[scale=2.6]{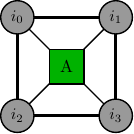}
\caption{A single tensor in the network. Its four tensor indices are labeled as $i_0i_1i_2i_3$. A grey circle represents a lattice site, or equivalently a tensor index. A green box represents a tensor.}
\label{fig:2dtn_right}
\end{figure}
One can also label the spin associated with tensor $A^a$ by its relative position, $x$, to $A^a$ with the notation $\sigma_x^a$.  
As shown in Fig. \ref{fig:2dtn_right}, there can be four relative positions in a 2D square lattice: $x = 0, 1, 2, 3$.  
Note that each spin in the 2D square lattice is associated with two tensors, and can serve, for example, both as $\sigma_0^a$ and $\sigma_3^b$ in Fig. \ref{fig:2dtn_left}.  
We have also defined the tensor indices of $A$ to be written as $A_{i_0i_1i_2i_3}$ where the tensor legs $0,1,2,3$ are labelled in Fig. \ref{fig:2dtn_right}. 
For example, to describe the homogeneous Ising model in Eq. \ref{eq:isingham}, the four-leg tensor $A^a$ has tensor elements: 
\begin{equation}
  \label{eq:A_tensor}
  A^a_{i_0i_1i_2i_3} = e^{K(\eta_{i_0} \eta_{i_1} + \eta_{i_1}\eta_{i_3} + \eta_{i_3} \eta_{i_2} + \eta_{i_2} \eta_{i_0})} , 
\end{equation}
where $\eta_i$ is the Ising spin associated with the tensor index $i$: 
\begin{equation}
  \label{eq:eta}
\eta_i \equiv \begin{cases}-1, &i = 0 \\ 1, &i=1\end{cases}. 
\end{equation} 

To perform MCRG, one needs to write the partition function as a configuration sum in terms of a Hamiltonian $H(\bm\sigma)$: 
\begin{equation}
  \label{eq:Z}
  Z = \sum_{\{\bm\sigma\}} e^{-H(\bm\sigma)}, 
\end{equation}
and the Hamiltonian needs to be written as a sum of $N_c$ coupling terms, $S_\beta(\bm\sigma)$:     
\begin{equation}
  \label{eq:Hamiltonian}
  H(\bm\sigma) = \sum_{\beta = 1}^{N_c} K_\beta S_\beta(\bm\sigma), 
\end{equation}
where $K_\beta$ is the coupling constant of the corresponding coupling term labeled by $\beta$. 
Traditionally, the coupling terms have been chosen as spin products, such as the nearest-neighbor product. 
The partition function in Eq. \ref{eq:Z_tensor} and Eq. \ref{eq:Z} will be equal if the Hamiltonian is given by: 
\begin{equation}
  H(\bm\sigma) = \sum_{a=1}^{N_A} \ln(A^a_{i_0 i_1 i_2 i_3}), 
\end{equation}
when $\sigma_0^a = \eta_{i_0}, \sigma_1^a = \eta_{i_1}, \sigma_2^a = \eta_{i_2}, \sigma_3^a = \eta_{i_3}$.
Here $N_A$ is the number of tensors in the network. 
Thus, the Hamiltonian which gives the same partition function as does the tensor network is the following:    
\begin{equation}
  H(\bm\sigma) = \sum_a^{N_A} \sum_{i_0i_1i_2i_3} \ln A^a_{i_0i_1i_2i_3} \delta_{\sigma_0^a,\eta_{i_0}} \delta_{\sigma_1^a,\eta_{i_1}} \delta_{\sigma_2^a,\eta_{i_2}}\delta_{\sigma_3^a,\eta_{i_3}}.
\end{equation}
In translationaly invariant systems, $\ln A_{i_0i_1i_2i_3}^a$ is independent from $a$, and 
\begin{equation}
  \label{eq:ks}
  \begin{split}
    H(\bm\sigma) &= \sum_{i_0i_1i_2i_3} \ln A_{i_0i_1i_2i_3} \sum_{a=1}^{N_A} \delta_{\sigma_0^a,\eta_{i_0}} \delta_{\sigma_1^a,\eta_{i_1}} \delta_{\sigma_2^a,\eta_{i_2}}\delta_{\sigma_3^a,\eta_{i_3}} \\ 
    &\equiv \sum_{\beta=1}^{N_c} K_\beta S_\beta(\bm\sigma).  
\end{split} 
\end{equation}
where we have identified the logarithm of each tensor element, $\ln A_{i_0i_1i_2i_3}$, as one coupling constant $K_\beta$ with a corresponding coupling term $S_\beta(\bm\sigma)$.
Thus, the ordered tuple $i_0i_1i_2i_3$ plays the role of $\beta$:    
\begin{equation}
  K_\beta = K_{i_0i_1i_2i_3}  = \ln A_{i_0i_1i_2i_3}
\end{equation}
and 
\begin{equation}
  S_\beta(\bm\sigma) = S_{i_0i_1i_2i_3}(\bm\sigma) = \sum_{a=1}^{N_A} \delta_{\sigma_0^a,\eta_{i_0}} \delta_{\sigma_1^a,\eta_{i_1}} \delta_{\sigma_2^a,\eta_{i_2}}\delta_{\sigma_3^a,\eta_{i_3}}. 
\end{equation}

For a tensor network with $n$ legs and bond dimension $\chi$ on each leg, there are therefore $N_c = \chi^n$ coupling terms, for example 16 in the case of 2D square lattice Isng model. 
% This number can be reduced by taking account of the symmetry of the system, as we will explain later. 
MCRG can then be performed with Eq. \ref{eq:ks}. 
% Note that in most TNRG algorithms, with the exception of Non-negative Tensor Network Renormalization \cite{tnr+}, the tensor elements are not always non-negative along the RG flow, and thus cannot be interpretated as a Hamiltonian. 
Since we are interested in the tensors before any renormalization, which are element-wise positive, taking the logarithm does not pose a problem.
\section{The Tangent Space to the Critical Manifold}
\label{sec:cmts}
When used to represent Hamiltonians with Eq. \ref{eq:ks}, the tensor network states in Fig. \ref{fig:2dtn_left} span a $k$ dimensional vector space of the Hamiltonians.
For example,  after taking account of the symmetry and the multiplicative freedom of a tensor network, $k = 3$ in the case of 2D square lattice Ising model as will be explained later.  
Embedded in this vector space is a $k-p$ dimensional manifold of the Hamiltonians that are critical, where $p$ is the number relevant RG operators of the sytem, or the codimension of the critical manifold.  
For the Ising models, for example, $p = 1$ in the $Z_2$ symmetric coupling space. 
Here we compute the tangent space to the critical manifold in this vector space exploiting the fact that the fixed-point Hamiltonian of the system after many iterations of majority-rule coarse-graining is invariant with respect to the change of the microscopic Hamiltonian along the critical manifold.   

In an MCRG calcuation, block-spins $\bm\sigma'$ are defined with a conditional probability $T(\bm\sigma' | \bm\sigma)$ given the unrenormalized spin configuration $\bm\sigma$, which realizes the scale transformation. 
In the rest of the paper,  we use $T$ to represent the $b=2$ majority-rule with a random pick on ties, where $b$ is the linear size of a spin block. 
In this coarse-graining procedure, the lattice is partitioned into blocks with size $b^d$, where $d$ is the space dimension. 
A coarse-grained spin $\sigma'$ is assigned to the block if among the spin values taken by the original spins $\bm\sigma$ in the block, $\sigma'$ is the most numerous.   
If, however, there are $m$ equally most numerous spins in a block where $m > 1$, i.e. a tie, the coarse-grained spin assigned to the block is chosen from these $m$ spins with even probability, i.e. $\frac{1}{m}$.   
This $b=2$ coarse-graining procedure can be iterated $n$ times to define the $n$th order block-spins $\bm\mu$, corresponding to a length transformation with scale factor $2^n$.    
The conditional probability $T^{(n)}(\bm\mu \ \bm\sigma)$ which describes this coarse-graining procedure is given by the $n$th composition of $T$:   
\begin{equation}
  T^{(n)}(\bm\mu|\bm\sigma) = \sum_{\bm\sigma^{(n-1)}}..\sum_{\bm\sigma^{(1)}} T(\bm\mu|\bm\sigma^{(n-1)}) \cdots T(\bm\sigma^{(1)}|\bm\sigma). 
\end{equation}
This defines the $n$th level renormalized Hamiltonian $H^{(n)}(\bm\mu)$ up to an $\bm\mu$-independent constant $g$:  
  \begin{equation}
  \label{eq:rg}
  \begin{split}
  H^{(n)}(\bm\mu) &\equiv -\ln \sum_{\bm\sigma} T^{(n)}(\bm\mu| \bm\sigma) e^{-H^{(0)}(\bm\sigma)} + g  
\\ 
  &= \sum_{\alpha} K^{(n)}_\alpha S_\alpha(\bm \mu) + g. 
\end{split} 
\end{equation}
where $K_\alpha^{(n)}$ is the $n$th level renormalized coupling constant associated with the coupling term $S_\alpha(\bm\mu)$. 
The constant $g$ corresponds to the background free energy of an RG transformation \cite{nauenberg}. 
It does not describe any nontrivial physics of the renormalized system, and is dropped at each RG iteration. 
The superscript $(0)$ refers to the unrenormalized system. 
As the coarse-graining is iterated, the change of $H^{(0)}$ along the critical manifold produces a change in $H^{(n)}$ that decays exponentially with $n$, and thus for sufficiently large $n$, $H^{(n)}$ can be viewed as invariant as $H^{(0)}$ changes along the critical manifold.    
In particular, up to exponentially small error in $n$, the tangent space to the critical manifold, will be the kernel of the Jacobian matrix of the $n$th level RG transformation:  
\begin{equation}
  \label{eq:matrix_A}
  \mathcal A_{\alpha\beta}^{(n, 0)} \delta K_\beta^{(0)} = 0, \hspace{5mm} \text{where }
  \mathcal A_{\alpha\beta}^{(n, 0)} = \frac{\partial K_\alpha^{(n)}}{\partial K_\beta^{(0)}}, 
\end{equation}
where the vector $\delta K^{(0)}_\beta$ is in TNCM.  
In general, the number of coupling terms to completely describe the renormalized Hamiltonian grows combinatorically with the renormalized lattice size and is infinite on an infinite lattice.
One cannot include all the coupling terms necessary to represent $H^{(n)}$ in practice, thus a finite number of renormalized coupling terms are used and a truncation error is incurred. 
However, as long as the truncation is well-defined, i.e. the truncation from $H^{(n)}$ to $K^{(n)}_\text{truncate}$ is unique, the invariance of $H^{(n)}$ with respect to the change of $K^{(0)}$ along the critical manifold dictates the invaraince of $K^{(n)}_\text{truncate}$ as well.    
Thus, one can replace the $K_\alpha^{(n)}$ in Eq. \ref{eq:matrix_A} with $K^{(n)}_{\alpha, \text{truncate}}$, and the kernel of $\mathcal A^{(n,0)}_{\alpha\beta}$ will still be the tangent space to the critical manifold with no truncation error. 
In the following, we use $K_\alpha^{(n)}$ to denote the truncated renormalized constants. 

Here we adopt the truncation scheme given by the Variational Monte Carlo Renormalization Group (VMCRG) \cite{vmcrg} which considers the bias potentials $V(\bm\mu)$ of the coarse-grained variable $\bm\mu$ expanded in a finite set of renormalized couplings $S_\alpha(\bm\mu)$ with parameters $J_\alpha$:     
\begin{equation}
  V(\bm\mu) = \sum_{\alpha = 1}^{N_c} J_\alpha S_\alpha(\bm\mu). 
\end{equation}
VMCRG minimizes a functional $\Omega[V]$ of the bias potentials $V$, defined as the the relative entropy 
\begin{equation}
  \Omega[V] \equiv D_{KL} (P_t || P_V) 
\end{equation}
between the bias distribution under $V$:   
\begin{equation}
  \label{eq:bias_ensemble}
  P_V(\bm\mu) \propto \sum_{\bm\sigma} \exp(-H^{(0)}(\bm\sigma)) T^{(n)}(\bm\mu | \bm\sigma) \exp(-V(\bm\mu)) 
\end{equation}
and the trivial distribution $P_t(\bm\mu)$. 
$\Omega[V]$ can also be written as \cite{vmcrg} 
\begin{equation}
  \Omega[V] = \ln \sum_{\bm\mu} e^{-H^{(n)}(\bm\mu) - V(\bm\mu)} + \sum_{\bm\mu} P_t(\bm\mu) V(\bm\mu) 
\end{equation}
up to a constant. 
It was shown that $\Omega[V]$ is convex \cite{vmcrg}. 
Writing the variational parameters of $V(\bm\mu)$ collectively as $\vec J = \{J_1, ..., J_\alpha, ..., J_{N_c}\}$, $\Omega(\vec J)$ can be viewed as a convex function with arguments $\vec J$. 
Its unique minimizer $V_\min = \sum_{\alpha} J_{\alpha,\min} S_\alpha(\bm\mu)$ can then be found using stochastic gradient descent with the gradient and Hessian of $\Omega(\vec J)$:
\begin{equation}
\label{eq:gradient}
\frac{\partial \Omega(\vec J)}{\partial J_\alpha} = - \braket{S_\alpha(\bm \mu)}_{V_{\vec J}} + \braket{S_\alpha(\bm \mu)}_{p_t}, 
\end{equation}
\begin{equation}
\label{eq:hessian}
\frac{\partial^2 \Omega(\vec J)}{\partial J_\alpha \partial J_\beta} = \braket{S_\alpha(\bm \mu) S_\beta(\bm \mu)}_{V_{\vec J}} - \braket{S_\alpha(\bm \mu)}_{V_{\vec J}}\braket{S_\beta(\bm \mu)}_{V_{\vec J}}. 
\end{equation}
Here $\braket{\cdot}_{V_{\vec J}}$ is the biased ensemble average under $V_{\vec J}$ and $\braket{\cdot}_{p_t}$ is the ensemble average under the target probability distribution $p_t$. 

If the set of coupling terms $S_\alpha$ is complete, $V_\min(\bm\mu) = \sum_{\alpha}J_{\alpha,\min}S_\alpha(\bm\mu) = - H^{(n)}(\bm\mu)$. 
We thus identify 
\begin{equation}
K_\alpha^{(n)} = -J_{\alpha,\min}. 
\label{eq:truncation}
\end{equation}
With a finite number of coupling terms, $J_{\alpha,\min}$ still exists but will not equal to the exact renormalized constants.
The truncated coupling constants are then defined by Eq. \ref{eq:truncation}. 
Within VMCRG, the truncated RG Jacobian matrix can be computed by inverting the following equation \cite{vmcrg}: 
\begin{equation}
  \label{eq:linear}
  \sum_\alpha \bbraket{S_\gamma(\bm\mu)}{S_\alpha(\bm\mu)}_{V_\min} \frac{\partial K_\alpha^{(n)}}{\partial K_\beta^{(0)}} = \bbraket{S_\gamma(\bm\mu)}{S_\beta(\bm\sigma)}_{V_\min}, 
\end{equation}
where $\bbraket{X}{Y}_{V_\min} \equiv \braket{XY}_{V_\min} - \braket{X}_{V_\min}\braket{Y}_{V_\min}$ is the connected correlation function of observable $X$ and $Y$ in the ensemble sampled according to Eq. \ref{eq:bias_ensemble} with $V_\min$.   
When the Monte Carlo is run with the bias potential $V_{\min}$, the correlation length and time of the bias distribution Eq. \ref{eq:bias_ensemble} will be greatly reduced and the sampling of the connected correlation functions much more efficient than the unbiased sampling.   

For systems whose critical manifold has codimension 1, its tangent space $\{\delta K_\beta^{0}\}$ is a hypersurface and is determined by its normal vector $\vec v$:    
\begin{equation}
  \sum_\beta v_\beta  \delta K^{(0)}_\beta = 0. 
\end{equation}
Thus each row vector of $A_{\alpha\beta}^{(n,0)}$ is a normal vector, and they should all be the same after normalization.  
If this is the case in our calculation, then it is an attestment to the assumption that an invariant fixed-point Hamiltonian exists under the majority-rule, that the microscopic Hamiltonian sampled is critical, and that the critical manifold has codimension 1.    
This will serve as a good consistent check for our calculation.  
\section{Numerical Results}
\label{sec:result}
\subsection{2D square lattice Ising model}
As demonstrated in Eq. \ref{eq:ks}, there is one coupling term for each tensor element.  
The space of Hamiltonians representable by a tensor network in Fig. \ref{fig:2dtn_left} is thus spanned by 16 coupling terms for the 2D square lattice Ising model.   
However, if we are only interested in the tensor networks representing the Hamiltonians symmetric under spin flip and the symmetry transformation of the underlying lattice, certain tensor elements should be restrained to equal one another, and there are only four truly distinct tensor elements, listed in Table \ref{table:ising_symmetry}.     
\begin{table}[!htb]
  \setlength{\tabcolsep}{0.8em}
\centering
  \begin{tabular}{ll} 
    \hline
    \hline
    $\beta$ & $i_0i_1i_2i_3$ \\ 
    \hline
    1 & 0000, 1111 \\
    2 & 0100, 0010, 0001, 0111, 1011, 1101, 1110, 1000 \\
    3 & 0110, 1001 \\ 
    4 & 0101, 0011, 1010, 1100\\ 
    \hline
    \hline
  \end{tabular}
  \caption{The tensor elements which are related to one another by symmetry.} \label{table:ising_symmetry}
\end{table}
Accordingly there are also only four couplings terms.  
The $\beta = 1$ coupling term, for example, will be defined as the sum of $i_0i_1i_2i_3 = 0000$ and $1111$ coupling terms:  
\begin{equation}
  \begin{split}
  S_1(\bm\sigma) &= \sum_{a = 1}^{N_A} \delta_{\sigma_0^a,\eta_0} \delta_{\sigma_1^a,\eta_0} \delta_{\sigma_2^a,\eta_0}\delta_{\sigma_3^a,\eta_0} \\ 
  &\hspace{10mm} + \sum_{a=1}^{N_A} \delta_{\sigma_0^a,\eta_1} \delta_{\sigma_1^a,\eta_1} \delta_{\sigma_2^a,\eta_1}\delta_{\sigma_3^a,\eta_1} 
\end{split}
\end{equation}
and 
\begin{equation}
  K_1 = \ln A_{0000} = \ln A_{1111}. 
\end{equation}
The coupling terms for $\beta = 2, 3, 4$ are analagously sumed with their respective symmetry partners according to Table \ref{table:ising_symmetry}.  
The four coupling terms thus formed, however, are not linearly independent, as evidenced by the equation
\begin{equation}
  \sum_{\beta=1}^{N_c = 4} S_\beta(\bm\mu) = N_A. 
\end{equation}
Here we identify two Hamiltonians if they are different only by a constant, so the constant function should be treated as the zero element of the vector space of Hamiltonians. 
The vector space of Hamiltonians we will consider is therefore only three dimensional: 
\begin{equation}
  \label{eq:expansion}
  \begin{split}
  H(\bm\mu) &= \sum_{\beta=1}^3 K_\beta S_\beta(\bm\mu) + K_4(N_A - S_1(\bm\mu) - S_2(\bm\mu) - S_3(\bm\mu)) 
  \\
  &= \sum_{\beta=1}^3 (K_\beta - K_4) S_\beta(\bm\mu) + \text{constant} 
  \\
  &= \sum_{\beta=1}^3 K'_\beta S_\beta(\bm\mu) + \text{constant}. 
\end{split} 
\end{equation}
The Jacobian matrix of the RG transformation which we will compute will be that of $K'_\beta$:
\begin{equation}
  \mathcal A^{(n,0)}_{\alpha\beta} = \frac{\partial K^{\prime (n)}_\alpha}{\partial K_\beta^{\prime (0)}}. 
\end{equation}

In Table \ref{tab:cmts2dising}, we report the matrix $\mathcal P_{\alpha\beta} = \frac{\mathcal A_{\alpha\beta}^{(n,0)}}{\mathcal A_{\alpha 1}^{(n,0)}}$, computed at the nearest-neighbor critical tensor in Eq. \ref{eq:A_tensor} with $K = 0.4406868$.     
Its rows are the normal vector to the tangent plane of TNCM, normalized so that the first element of the vector is 1.     
\begin{table}[htb!]
  \setlength{\tabcolsep}{0.8em}
\centering
  \begin{tabular}{llll} 
    \hline
    \hline
    $\alpha$ & $\mathcal P_{\alpha 1}$ & $\mathcal P_{\alpha 2}$ & $\mathcal P_{\alpha 3}$\\  
    \hline
    1 & 1 & -0.522(1) & -0.0184(3) \\ 
    2 & 1 & -0.522(7) & -0.018(1)  \\ 
    3 & 1 & -0.522(3) & -0.0185(3) \\
    \hline
    \hline
  \end{tabular}
  \caption{The matrix $\mathcal P_{\alpha\beta} = \frac{\mathcal A_{\alpha\beta}^{(n,0)}}{\mathcal A_{\alpha 1}^{(n,0)}}$ for the isotropic 2D square Ising model. 
  A $256^2$ lattice was used with the renormalization level $n = 5$. 
  The simulations were performed on 16 cores independently, each of which ran $3\times 10^6$ Metropolis MC sweeps.    
  The mean is cited as the result and twice the standard error as the statistical uncertainty.  
  }
\label{tab:cmts2dising}
\end{table}
The consistency among the different rows confirms our assumptions. 
The statistical uncertainties of the result, however, are different for different rows, because the connected correlation functions of different coupling terms $\alpha,\beta$ have different variance in an MC sampling.    
One should always cite the result with the least statistical uncertainty. 
In converting the computed $\delta K'_\beta$ with $\beta = 1,2,3$ to the actual change in the tensor elements, $\delta K_\beta$ with $\beta = 1,2,3,4$, one is free to choose the values of $\delta K_\beta$ so long as the resultant $\delta K'_\beta = \delta K_\beta - \delta K_4$ conforms to the computed value. 
Here we take $\delta K_\beta = \delta K'_\beta$ for $\beta = 1,2,3$ and $\delta K_4 = 0$. 
This freedom is the same multiplicative normalization freedom in a tensor network state.    

In the end, we present the tangent space to TNCM in matrix form by combining $i_0i_1$ of $A_{i_0i_1i_2i_3}$ as a row index $m = i_0 + 2i_1$ and $i_2i_3$ as a column index $n = i_2 + 2i_3$.  
To the linear order of $\delta K_2$ and $\delta K_3$, the set of all critical Hamiltonians representable by of a tensor network in Fig. \ref{fig:2dtn_left} that respect the symmetry of the 2D square lattice is
\begin{widetext}
\begin{equation}
  \label{eq:ising_matrix}
  \ln A_{i_0i_1,i_2i_3} 
= K_c 
\begin{pmatrix}  
  4 & 0 & 0 & 0 \\
  0 & 0 & -4 & 0 \\
  0 & -4 & 0 & 0 \\
  0 & 0  & 0 & 4
\end{pmatrix} 
+ \delta K_2 
\begin{pmatrix}  
  0.522(1) & 1 & 1 & 0 \\
  1 & 0 & 0 & 1 \\
  1 & 0 & 0 & 1 \\
  0 & 1 & 1 & 0.522(1) 
\end{pmatrix} 
+ \delta K_3 
\begin{pmatrix}  
  0.0184(3) & 0 & 0 & 0 \\
  0 & 0 & 1 & 0 \\
  0 & 1 & 0 & 0 \\
  0 & 0 & 0 & 0.0184(3) 
\end{pmatrix}.  
\end{equation}
\end{widetext}
where $K_c = 0.4406868$, and $\delta K_2$ and $\delta K_3$ are infinitesimally small, but otherwise arbitrary.  
\subsection{2D square lattice three-state Potts mode}
Next consider the three-state Potts model on a 2D square lattice: 
\begin{equation}
  \label{eq:3potts}
  H(\bm\sigma) = -K \sum_{\braket{x,y}} \delta_{\sigma_x\sigma_y}, 
\end{equation}
where $\braket{x,y}$ denotes nearest-neighbor pairs and $K > 0$. 
The spin at each lattice site takes on $\sigma_x = 0, 1, 2$ three possible values.  
The system experiences a continuous phase transition at $K_c = 1.005053...$ \cite{potts}.
This model is also representable by a tensor network in Fig. \ref{fig:2dtn_left} with bond dimension $\chi = 3$. 
The map from tensor indices to spin variables is simply the identity map:   
\begin{equation}
  \eta_i = i, \text{ for }i = 0, 1, 2. 
\end{equation}
The tensor-representable Hamiltonian can again be written as in Eq. \ref{eq:ks} with $N_c = 3^4 = 81$. 

\begin{figure}[h]
\centering
  \centering
  \includegraphics[scale=0.38]{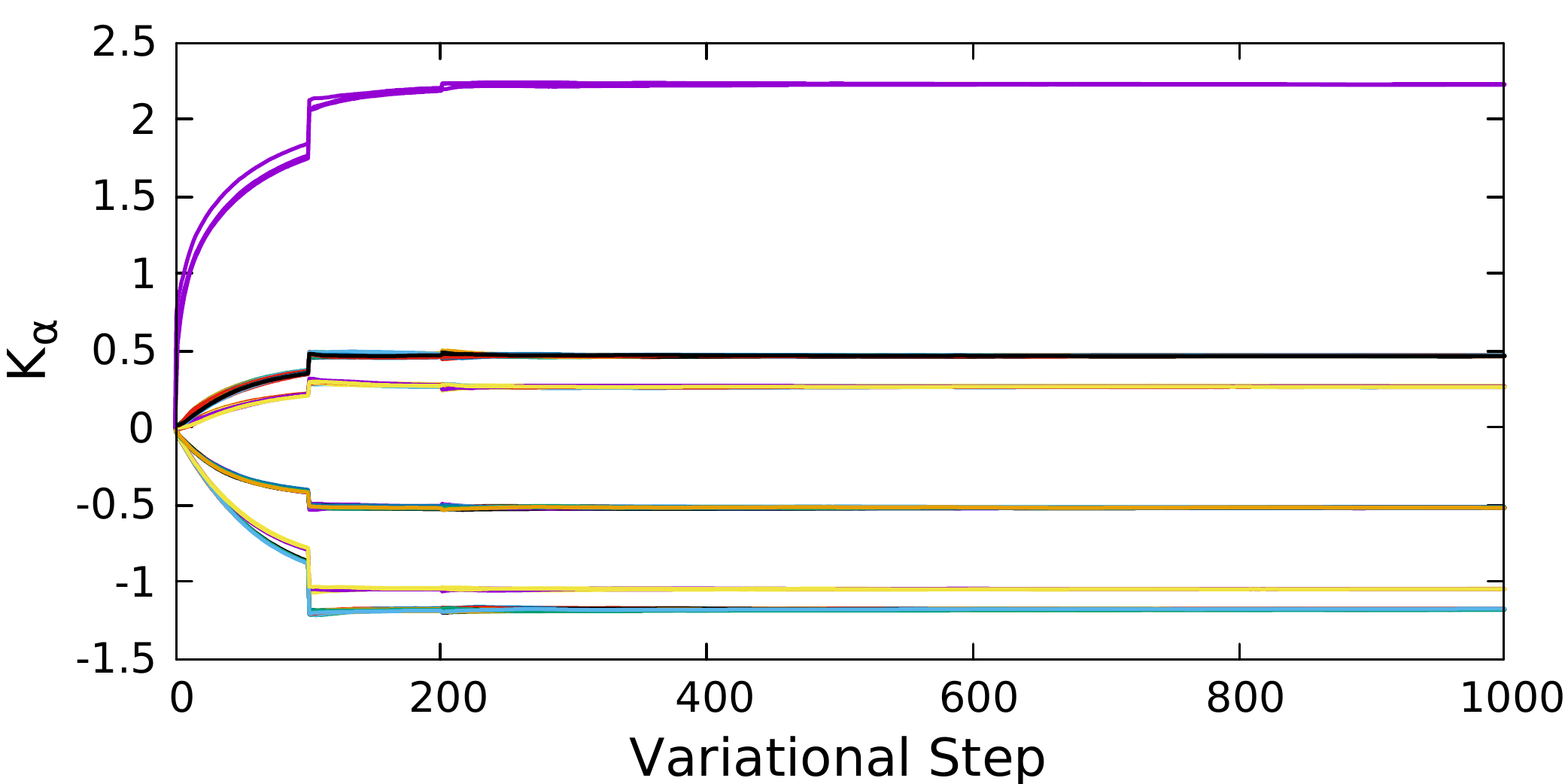}
\caption{Optimization trajectory of the tensor network renormalized constants for the three-state Potts model on a $16^2$ lattice at $K = 1.005053$. All $81$ renormalized constants are independently optimized and shown. 
Each surve represents one coupling term.}
\label{fig:vary_potts}
\end{figure}

Unlike the 2D Ising model, the symmetry classes of the coupling terms are onerous to identify by hand.   
VMCRG, however, can be used to find the symmetry partners of the many coupling terms.  
To perform this task, the renormalized constants after one iteration of $2\times 2$ majority-rule is determined with all of the 81 couplings terms, shown in Fig. \ref{fig:vary_potts}.  
The couplings with the same renormalized constants (up to some noise) are then the symmetry partners with one another.  
There are thus six symmetry classes and coupling terms, listed in Table \ref{tab:2dpotts_symm}.  
\begin{table}[htb!]
  \setlength{\tabcolsep}{0.8em}
\centering
  \begin{tabular}{ll} 
    \hline
    \hline
    $\beta$ & $i_0i_1i_2i_3$ \\ 
    \hline
    1 & 0000 \\
    2 & 1000\\
    3 & 1100\\ 
    4 & 2100\\ 
    5 & 0110\\ 
    6 & 2110\\ 
    \hline
    \hline
  \end{tabular}
  \caption{The tensor elements which belong to distinct symmetry classes.
  Only one representative of each class is listed. 
  }
  \label{tab:2dpotts_symm}
\end{table}

Eliminating the linear dependence, we use the first five coupling terms to span the space of Hamiltonians representable by a tensor network, in which is embedded a four-dimensional critical manifold. 
(The codimension of the the critical manifold for the 2D three-state Potts model is also one.)   
The tangent space to TNCM is again reported as the matrix $\mathcal P_{\alpha\beta} = \frac{\mathcal A_{\alpha\beta}^{(n,0)}}{\mathcal A_{\alpha 1}^{(n,0)}}$ in Table \ref{tab:cmts2dpotts}, from which its matrix form can be constructed as in Eq. \ref{eq:ising_matrix}.  
\begin{table}[htb!]
  \setlength{\tabcolsep}{0.8em}
\centering
  \begin{tabular}{lllll} 
    \hline
    \hline
    $\mathcal P_{\alpha 1}$ & $\mathcal P_{\alpha 2}$& $\mathcal P_{\alpha 3}$ & $\mathcal P_{\alpha 4}$ &$\mathcal P_{\alpha 5}$\\ 
    \hline
    1 & -0.381(2)& -0.363(1) & -0.216(1) & -0.0117(2)\\ 
    1 & -0.381(2)& -0.363(1) & -0.216(1) & -0.0118(3)\\ 
    1 & -0.378(3) & -0.364(2)  & -0.218(2) & -0.0123(6)\\
    1 & -0.382(7) & -0.361(4)  & -0.218(3) & -0.012(1) \\ 
    1 & -0.39(5)  & -0.36(4)   & -0.21(2)  & -0.015(6) \\ 
    \hline
    \hline
  \end{tabular}
  \caption{The matrix $\mathcal P_{\alpha\beta} = \frac{\mathcal A_{\alpha\beta}^{(n,0)}}{\mathcal A_{\alpha 1}^{(n,0)}}$ for the isotropic 2D square three-state Potts model. 
  A $256^2$ lattice was used with the renormalization level $n = 5$. 
  The simulations were performed on 16 cores independently, each of which ran $9\times 10^5$ Metropolis MC sweeps.    
  The mean is cited as the result and the standard error as the statistical uncertainty. 
  }
\label{tab:cmts2dpotts}
\end{table}
% The tangent space to TNCM is presented in matrix form in Appendix \ref{subsec:2dpotts_cmts}. 
\subsection{3D cubic lattice Ising model}
\begin{figure}[htb]
  \includegraphics[scale=0.7]{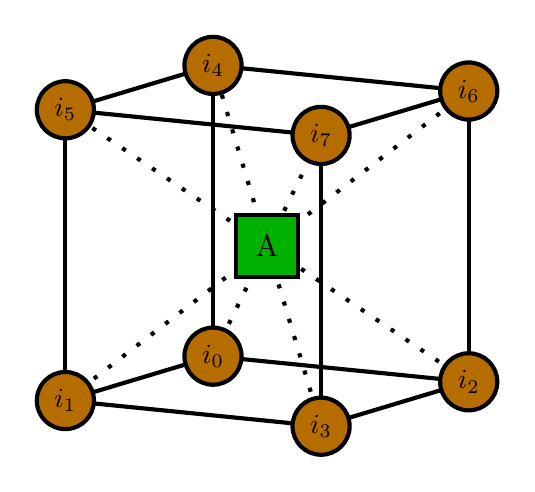}
  \caption{The tensor in cubic-lattice tensor network. It is associated with 8 spins.}
  \label{fig:3dtn}
\end{figure}

In the end, we consider the Ising model Hamiltonian of Eq. \ref{eq:isingham} in the 3D cubic lattice.
Although there has not been TNRG algorithms that generate a proper RG flow for this model, we still present here the result of TNCM in anticipation of further advancement of TNRG in 3D.   
In the cubic lattice, the tensors have eight legs, shown in Fig. \ref{fig:3dtn}.    
They are placed in a network where each spin is associated with two tensors and each nearest-neighbor bond of the lattice is accounted once by the network, similar to the case in two dimension (Fig. \ref{fig:2dtn_left}).  
$2^8 = 256$ coupling terms are present by Eq. \ref{eq:ks}.     
Among them are $13$ symmetrized coupling terms, found with VMCRG, listed in Table \ref{tab:3dising_symm}.  
\begin{table}[htb!]
  \setlength{\tabcolsep}{0.8em}
\centering
  \begin{tabular}{llll} 
    \hline
    \hline
    $\beta$ & $i_0i_1i_2i_3i_4i_5i_6i_7$ & $\beta$ & $i_0i_1i_2i_3i_4i_5i_6i_7$ \\ 
    \hline
    1 & 00000000 & 8 & 11101000\\
    2 & 10000000 & 9 & 10011000\\
    3 & 11000000 & 10 & 11011000\\ 
    4 & 01100000 & 11 & 01111000\\ 
    5 & 11100000 & 12 & 00111100\\ 
    6 & 11110000 & 13 & 10010110\\ 
    7 & 01101000 \\
    \hline
    \hline
  \end{tabular}
  \caption{The tensor elements which belong to distinct symmetry classes.
  Only one representative of each class is listed. 
  }
  \label{tab:3dising_symm}
\end{table}
Again, eliminating the linear dependence, the first 12 coupling terms are used to span the vector space of Hamiltonians representable by a 3D tensor network, which admits a $11$-dimensional critical manifold.  
The matrix $\mathcal P_{\alpha\beta} = \frac{\mathcal A_{\alpha\beta}^{(n,0)}}{\mathcal A_{\alpha,1}^{(n,0)}}$ is rather large, so we only cite here the row with the least statistical uncertainty in Eq. \ref{eq:3dising_cmts}, and note that the consistency among the rows is indeed observed.  
A $64^3$ lattice was used with the renormalization level $n = 3$. 
The simulations were performed on 464 cores independently, each of which ran $4.9\times 10^5$ Metropolis MC sweeps.    
The mean is cited as the result and twice the standard error as the statistical uncertainty.  
  \begin{equation}
  \label{eq:3dising_cmts}
  \begin{split}
    \mathcal P_{1\beta} =  
  \big[1, \,\, &0.590(4), \,\, -0.151(3), \,\, -0.037(2), \,\, -0.621(3)
  \\
  -&0.164(2), \,\, -0.0241(8), \,\, -0.086(2), \,\, -0.195(2), \,\, 
  \\
  -&0.218(2), \,\, -0.076(1), \,\, -0.0176(6)  \big].  
\end{split}
\end{equation}

\section{Conclusion}
\label{sec:discussion}
In this paper, we have shown how MCRG can be performed with coupling constants encoded in a tensor network.  
While the associated finite number of coupling terms do not represent the exact renormalized Hamiltonian, the tangent space to the critical manifold can still be obtained free of truncation error. 
The tangent spaces to TNCM are computed for three example models with MCRG.    
With this knowledge, the requirement B mentioned in Sec. \ref{sec:intro} can then be checked for the various TNRG algorithms to achieve a further understanding of how irrelevant operators get suppressed in these algorithms.       
\begin{acknowledgments}
All the codes used in this project were written in C\texttt{++}, and will be available upon request. The author acknowledges instruction from his advisor Roberto Car and support from DOE Award DE-SC0017865. He is also thankful for a critical reading of the manuscript by Linfeng Zhang.  
\end{acknowledgments}

\bibliographystyle{apsrev}
% \bibliography{abc}

\end{document}